\documentclass[aps,twocolumn,showpacs,nofootinbib,showkeys,superscriptaddress,preprintnumbers,longbibliography,10pt]{revtex4-2}
\usepackage[utf8]{inputenc}

\usepackage{latexsym}
\usepackage{epsfig}
\usepackage{graphicx,subfigure}
\usepackage[normalem]{ulem}
\usepackage{color}
\usepackage{xcolor}
\usepackage{multirow}
\usepackage{hyperref}
\usepackage{numprint}
\usepackage{eucal}
\usepackage{amsmath}
\usepackage{amssymb}
\usepackage{amsmath,amssymb} 
\usepackage{wasysym}

\usepackage{soul}

\DeclareFontFamily{U}{mathb}{\hyphenchar\font45}
\DeclareFontShape{U}{mathb}{m}{n}{
      <5> <6> <7> <8> <9> <10> gen * mathb
      <10.95> mathb10 <12> <14.4> <17.28> <20.74> <24.88> mathb12
      }{}

\definecolor{blue}{rgb}{0,120,250}

\usepackage{xcolor}

\newcommand{\ssout}[1]{}
\begin{document}

\def\ga{\mathrel{\raise.3ex\hbox{$>$\kern-.75em\lower1ex\hbox{$\sim$}}}}
\def\la{\mathrel{\raise.3ex\hbox{$<$\kern-.75em\lower1ex\hbox{$\sim$}}}}

\def\be{\begin{equation}}
\def\ee{\end{equation}}
\def\bea{\begin{eqnarray}}
\def\eea{\end{eqnarray}}

\def\betap{\tilde\beta}
\def\del{\delta_{\rm PBH}^{\rm local}}
\def\Msun{M_\odot}

\newcommand{\dd}{\mathrm{d}} % For the derivatives
\newcommand{\Mpl}{M_P} % For the reduced Planck mass
\newcommand{\mpl}{m_\mathrm{pl}} % For the Planck mass

\definecolor{ao}{rgb}{0.0, 0.0, 1.0}
 
\newcommand{\CHECK}[1]{{\color{red}~\textsf{#1}}}

\title{Can Primordial Black Holes form in the Standard Model ?} 

\author{Ioanna Stamou}
\affiliation{Service de Physique Th\'eorique, Universit\'e Libre de Bruxelles (ULB), Boulevard du Triomphe, CP225, B-1050 Brussels, Belgium}

\author{Sebastien Clesse}
\affiliation{Service de Physique Th\'eorique, Universit\'e Libre de Bruxelles (ULB), Boulevard du Triomphe, CP225, B-1050 Brussels, Belgium}

%\date{\today}
\begin{abstract} 
We investigate the viability of primordial black hole (PBH) formation in the Standard Model (SM) in a scenario that does not rely on specific inflationary features or any exotic physics such as phase transitions or non-minimal coupling to gravity.  If the Brout-Englert-Higgs (BEH) field lies exactly at the transition between metastability and stability, its potential exhibits an inflexion point due to radiative corrections.  
The BEH can act like a stochastic curvaton field, leading to a non-Gaussian tail of large curvature fluctuations that later collapse into PBHs when they re-enter inside the horizon.  This scenario would require a precise value of the top-quark mass to ensure the Higgs stability, which is disfavored but still consistent with the most recent measurements.
%and testable with future collider data.  
However, we also find that large curvature fluctuations are also generated on cosmological scales that are inconsistent with cosmic microwave background (CMB) observations.  We therefore conclude that the SM cannot have led to the formation of PBHs based on this mechanism.  Nevertheless, a variation of the scenario based on the Palatini formulation of gravity may have provided the conditions to produce stellar-mass PBHs with an abundance comparable to dark matter, without producing too large curvature fluctuations on cosmological scales.     
\end{abstract}
%\pacs{04.70.Bw, 97.60.Lf, 95.35.+d}

\maketitle

\textbf{Introduction --} Primordial black holes (PBHs) have recently been promoted among the hottest topics in Cosmology. They could explain a sizeable or even the totality of the dark matter (DM) in the Universe as well as some intriguing properties of black hole binary mergers~\cite{Carr:2023tpt} observed by the LIGO-Virgo-Kagra (LVK) collaboration~\cite{Abbott:2016blz,LIGOScientific:2016dsl,LIGOScientific:2018mvr,LIGOScientific:2020ibl,LIGOScientific:2021usb,LIGOScientific:2021djp}, such as black holes in the pair-instability mass gap heavier than $60 M_\odot$ like in GW190521 \cite{LIGOScientific:2020iuh}, or binaries with low-mass ratios like GW190814 \cite{LIGOScientific:2020zkf}.  Other possible clues for the existence of PBHs~\cite{Clesse:2017bsw,Carr:2019kxo,Cappelluti:2021usg,Carr:2023tpt} include  microlensing observations of quasars or stars, spatial correlations in the source-subtracted X-ray and infrared background radiations~\cite{Kashlinsky:2016sdv}, the existence of super-massive black holes very early in the Universe history or unexpected galaxies at high redshifts recently observed by the James Webb Space Telescope \cite{Yuan:2023bvh,Hutsi:2022fzw}.   At the same time, the abundance of PBHs is subject to stringent limits  from various probes, see e.g.\cite{Carr:2020xqk,LISACosmologyWorkingGroup:2023njw}  for recent reviews.   There are only two possible mass ranges for PBHs to significantly contribute to the DM:  the asteroid-mass region between $ 10^{-16} M_\odot $ and $10^{-11} M_\odot$ and the solar-mass region that is only mildly constrained by microlensing surveys.

PBHs may have collapsed in the very early Universe~\cite{Hawking:1971ei,Carr:1974nx} from pre-existing inhomogeneities on smaller scales than the ones probed by the cosmic microwave background (CMB) anisotropies and the large scale structures of the Universe.  These fluctuations can typically be produced during inflation but other mechanisms are also possible (see \cite{Ozsoy:2023ryl,LISACosmologyWorkingGroup:2023njw,Khlopov:2008qy}  for reviews).  However, despite their interest on an observational point-of-view, almost all PBH models suffer from severe fine-tuning and coincidence issues~\cite{Cole:2023wyx}, e.g. in order to explain a comparable abundance to baryons or masses comparable to stars.  In this context, a recent progress has been to identify the role of the QCD epoch in boosting the PBH formation \cite{Byrnes:2018clq,Carr:2019hud,Escriva:2022bwe,Franciolini:2022tfm}  and shaping their mass distribution with features induced by the transient variation of the equation-of-state of the Universe at this epoch.   This effect provides a natural explanation to a population of stellar-mass PBHs.  Nevertheless, PBH still remain a rather unnatural hypothesis and all the models proposed so far require exotic ingredients such as features in the scalar field potential responsible for cosmic inflation, a regime of ultra-slow-roll, multi-fields, a non-minimal coupling to gravity, etc, combined with unnatural parameter fine-tuning.  So the present situation on the theory side is still rather unsatisfactory.

In this \textit{letter}, we examine if PBHs could have formed in the framework of the Standard Model (SM), with the common addition of standard slow-roll inflation, without requiring any exotic ingredient. 
%than the usual phase of slow-roll inflation.  
The considered PBH formation mechanism relies on the Brout-Englert-Higgs (BEH) field that can be a light quantum stochastic spectator field during inflation.  For special values of the BEH mass, the top quark mass and the strong coupling constant, the radiative-corrected BEH potential is stable and exhibits an inflexion point.  We show that in the regions of the Universe where the BEH field after inflation end up to be very close to this inflexion point, curvature fluctuations are generated when the BEH density transiently dominates, in addition to the ones from slow-roll inflation.  These are generated on all scales and have a non-Gaussian tail on small scales, allowing large fluctuations to collapse into PBHs when they later re-enter inside the horizon.  We compute their statistics using the stochastic $\delta N$ formalism \cite{Lyth:2005fi,Langlois:2008vk}.
However, we also find that the regions where the BEH field can generate PBHs also produce too large curvature fluctuations on cosmological scales, which is ruled out by CMB observations.  Nevertheless we discuss the conditions under which such a scenario could have been viable and believe that the mechanism remains interesting in more particular scenarios.  As an example, we study the PBH formation from the BEH field in the Palatini formulation of Gravity \cite{Bauer:2008zj}, for which the CMB constraints can be satisfied.  

A connection between the BEH and PBH formation was already envisaged in different contexts.  One can mention the model of critical Higgs inflation \cite{Ezquiaga:2017fvi,Bezrukov:2017dyv}  where the BEH has a non-minimal coupling and plays the role of the inflaton along an inflexion-point potential.  Another possibility is that the BEH field is trapped in a false vacuum \cite{Maeso:2021xvl}.  Finally, curvature fluctuations from the BEH as a stochastic field was considered in \cite{Passaglia:2019ueo}, also finding that it is inconsistent with CMB observations, but their mechanism was different because curvature fluctuations were produced during inflation, and not after it as in our model.  

This \textit{letter} is organized as follows:  after introducing the BEH potential, we calculate the BEH probability distribution reached at the end of inflation, due to its stochastic dynamics.  Then we relate these field fluctuations to the generation of curvature fluctuations after inflation, which are compared to the CMB primordial power spectrum. Then we explore the PBH formation in the Palatini formulation of Gravity and we compute the expected PBH mass distribution from the statistics of the curvature fluctuations generated by the BEH field.  Finally we discuss our findings and present our conclusions and some perspectives. 

%Even more, this scenario does not require any tuning of parameters other than ones that are already observed in the Standard Model.  It still requires an usual phase of single-field slow-roll inflation but without imposing any strong condition on the inflaton itself.  The curvature fluctuations at the origin of PBH formation 

%______________________________________________
%_______________________________________________

\textbf{The BEH potential --}  In our scenario, we consider an inflaton field $\phi$ whereas the BEH field $\rm{h}$ is a light stochastic spectator field, minimally coupled to gravity, with a mass $\rm{m_{h} \ll H}$ during inflation, such that it typically experiences quantum fluctuations of order $\rm{H/2 \pi }$ over one e-fold of expansion.  We chose the top quark mass, the BEH bare mass and the QCD coupling constant such that the BEH potential exhibits an inflexion point.
The effective BEH potential 
at large field values, where $\mathrm{h \gg \upsilon}$, with $\mathrm{v}$ being the scalar vacuum expectation value (VEV), 
%the effective potential of the Standard Model Higgs 
is given by
\begin{equation}
    V(h)= \frac{\lambda(h)}{4}h^4,
    \label{fig:pot}
\end{equation}
where the self-coupling $\lambda(h)$ is determined by the so-called $\beta$ function, $\beta_{\lambda}={\rm d} \lambda / {\rm d} \ln \mu$.
The $\mathrm{\beta}$ function has been computed from the contributions of top Yukawa couplings at one-loop level and higher orders for which we also consider the strong couplings and the BEH self-coupling as well as the anomalous dimensions of the scalar field~\cite{Chetyrkin:2012rz, Bezrukov:2012sa}. 
%where  the running $\lambda(h)$ can been evaluated from $\beta$ function.
For this purpose, we 
%For the computation of $\mathrm{\lambda(h)}$, we 
have used the code provided in \cite{Bezrukov:2012sa}, based on \cite{Chetyrkin:2012rz, Bezrukov:2012sa}, that includes the two-loop and three-loop radiative corrections to the BEH potential.

\begin{figure}[h!]
\centering
\includegraphics[width=80mm]{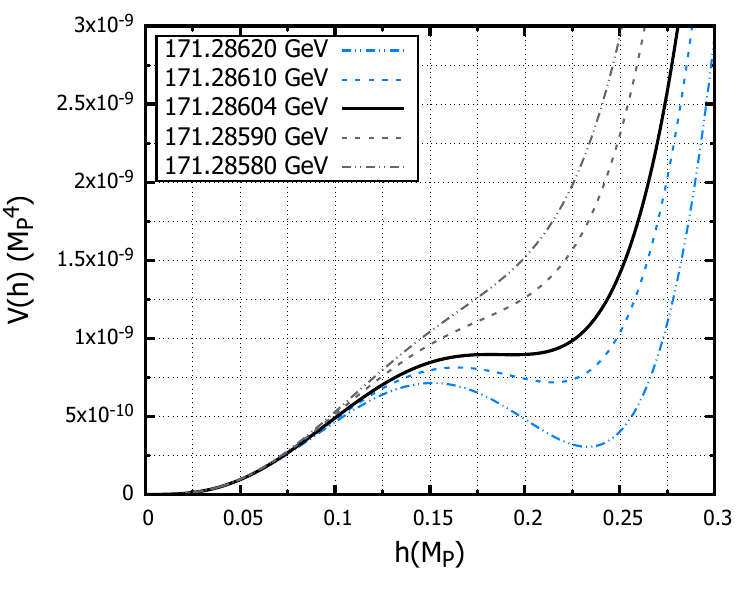}
\caption{The BEH field potential with a near-inflection point (black line). The dashed lines correspond to different values of the top  quark mass, given in the legend. $\mathrm{M_P}$ is the reduced Planck mass.  The mass of BEH is fixed to ${m_{\rm h}=125.7}$ GeV and the strong coupling constant to ${\alpha_{\rm s}=0.1184}$. }
\label{f1}
\end{figure}

The two important parameters in the SM that determine the EW phases are the BEH mass $\mathrm{m_h}= (124.94 \pm 0.17)$ GeV and the top-quark mass $\mathrm{m_t} = (171.1 \pm 0.4) $ GeV \cite{ATLAS:2022zjg,ATLAS:2019guf}. 
%It has been noticed that  
These measured values of $\mathrm{m_h}$ and $m_t$ place the SM almost near the transition between the stability and metastability regimes \cite{Degrassi:2012ry,Buttazzo:2013uya}. 

In Fig. \ref{f1} we depict the potential for different values of the top-quark mass, assuming the best fits of the BEH mass and strong coupling constant. 
%of the EW scales for the 
%Higgs 
%mass $\mathrm{m_h=125.7 }$ GeV and of the strong coupling $a_s=0.1184$. 
For a fine-tuned value of $m_t = 171.28604$ GeV, we
obtain a near-inflexion point in the scalar potential (black line).  
% [2]
\textcolor{black}{This inflection point marks the transition from the stable to the metastable regime. In Fig. 1, we illustrate how the top quark mass sensitivity can lead to this inflection point. For simplicity, we describe the scenarios depicted by gray lines as belonging to the stable regime and those by blue lines as metastable. It's important to note that all cases represented in the plot do not exhibit negative potential energy. }
%\textcolor{red}{The gray lines are for top quark masses in the stable regime of the SM vacuum, and the blue lines are for values in the metastable regime. }
%

%______________________________________________
%_______________________________________________

\textbf{Curvature fluctuations from the BEH stochastic spectator field  --} 
During inflation, the BEH field is typically very light with a mass $\rm{m_h \ll H}$, with $\rm{H}$ being the expansion rate.  In this limit, one can calculate the evolution, in e-fold time $N_{\rm inf}$, of the variance of its quantum fluctuations.  We distinguish between the field fluctuation in a Hubble-sized region at time $N_{\rm inf}$ with respect to the mean field value in the immediately surrounding region, $\delta h_{\rm in}$, and the fluctuation of that outside mean field value compared to the averaged field value in the whole observable patch of the Universe, denoted $\delta h_{\rm out}$. 
% [3B]
\textcolor{black}{ We assume that the first slow-roll parameter $\epsilon_1$, definied as $\epsilon_1 \equiv -{{\rm d} \ln H}/{{\rm d}N}$, is slowly varying during inflation  while the second parameter $\epsilon_2 \equiv {{\rm d} \ln \epsilon_1}/{{\rm d}N}$ is assumed to be constant.} One gets \cite{Stamou:2023vft}
%such that the variance of a quantum fluctuation $\delta_h_{\rm in}$ at the e-fold time $N_{\rm inf}$ (  evolve as (REF) %of the field $\mathrm{h}$ is given: 
\begin{equation}
    \langle \delta h_{\rm out}^2(N_{\rm inf}) \rangle \simeq\frac{{H}_{\rm *}^2}{8 \pi^2 \epsilon_{1*}}\left( 1-e^{-2\epsilon_{1*} (N_{\rm inf}-N_{\rm inf,*}) }\right)
\label{eq:field1}
\end{equation}
%and for $\mathrm{\langle \delta {h_{in}}(N_{inf}) \rangle}$  we have the following:
\begin{equation}
    \langle \delta h_{\rm in}^2(N_{\rm inf}) \rangle \simeq\frac{{H}^2_{*}}{4 \pi^2 } \exp\left\{ -2\frac{\epsilon_{1*}}{\epsilon_{2*}} \left[ e^{\epsilon_{2*} (N_{\rm inf}-N_{\rm inf,*}) } - 1 \right]\right\}, 
\label{eq:field2}
\end{equation}
where we assume that the slow-roll parameters at the time a pivot scale $k_*$ exited the horizon, are given by $\mathrm{\epsilon_{1*}}=0.0002$ and $\mathrm{\epsilon_{2*}}=0.035$. These values are compatible with the ones derived for $\mathrm{R^2}$ theories such as the Starobinsky inflation model, and with CMB observations. The  Hubble rate parameter at the same time is ${H_{*}}= 5.8 $ $\times10^{-6} M_{\rm P}$ in order to get the correct CMB normalization, while the tensor-to-scalar ratio is then $\mathrm{}r=16 \epsilon_{1*} = 0.0032$.  %\textcolor{ao}{ [3B]For more details about the methodology see Ref. \cite{Stamou:2023vft}.}  
The e-fold time during inflation $N_{\rm inf}$ is arbitrarily \textcolor{black}{scaled} such that \textcolor{black}{the e-fold at horizon exit of $k_*$ is} $N_{\rm inf *}  = 0$.  \textcolor{black}{We are interested in the evolution of field fluctuations at $N_{\rm inf} >0$ that are inside Universe patches of similar size than our observable Universe. }
% [3A]
\textcolor{black}{ We denote  $\rm{\langle h \rangle}$  the mean BEH field value in our observable patch and treat it as a free parameter.}  

 The equations governing the evolution of $h$ in the subsequent matter era are 
\begin{equation}
h '' +h' \frac{1}{2 \rho} \left(-3 \rho_{\rm m}e^{-3N}+h'\frac{\partial V}{\partial h}\right)+3h' +\frac{3M_{\rm P}^2}{\rho}\frac{\partial V}{\partial h}=0
\label{eq:field}
\end{equation}
where primes denotes derivatives in respect to e-folds, $M_{\rm P}$ is the reduced Planck mass and $\rho_{\rm m}$ is the matter density at some initial time before the BEH starts to dominate.  We thus assume that the BEH quickly dominates during the reheating phase, but alternatively, one may have considered a rapid transition to the radiation era and adapt the above equation accordingly, without significantly changing our results.
%[3A]
 \textcolor{black}{  In the following we define $\rm {h_{ic}}$ as the considered  initial field values for the numerical integration of Eq. \ref{eq:field} and $\rm{h_{cr}}$ as the field value at the inflection point (or local maximum). }
%We work  in Planck units, $\mathrm{M_P=1}$.  
The total energy density $\mathrm{\rho}$ is given by
\begin{equation}
\rho =\rho_{\rm m} e^{-3N} +V(h),
\end{equation}
and we consider an initial e-fold time $N=0$ for the spectator-field domnation phase that corresponds to $\rho_{\rm m} = 10 V(h)$ initially.
 %%[3E-3FF]
\textcolor{black}{ This choice ensures that the field evolution is frozen by the Hubble friction from the matter density before the numerical integration starts.  A larger initial value of $\rho_{\rm m}$ would not have led to sizeable curvature fluctuations.  We end the numerial integration when the first slow-roll paramter reaches unity and negelect the kinetic terms during the slow-roll phase as well as the subsequent oscillatory phase, because they do not induce sizeable additional curvature fluctuations, as discussed in~\cite{Stamou:2023vft}. }

\begin{figure}[t]
\centering
\includegraphics[width=80mm]{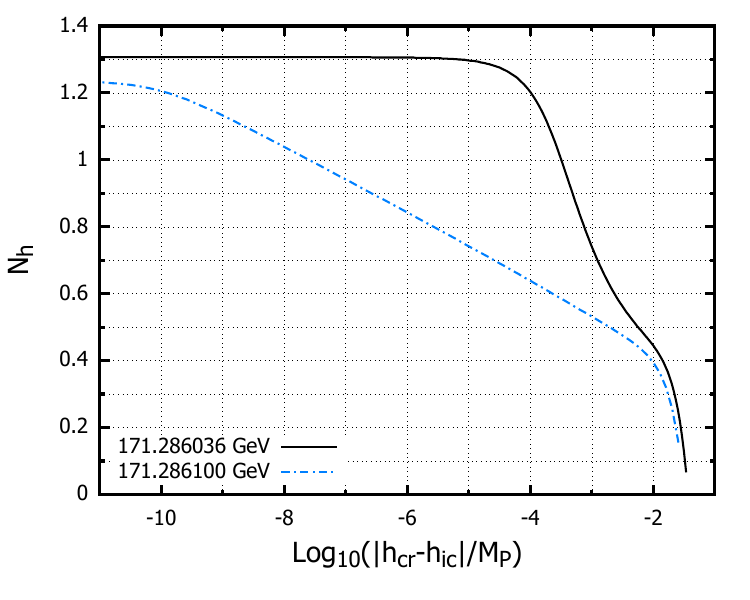}
\caption{Number of e-folds realised during the BEH spectator field domination phase, as a function of the difference between the initial BEH field value $h_{\rm ic}$ and the potential maximum $h_{\rm cr}$, for top quark masses corresponding to the metastable regime, with the BEH mass and the strong coupling constant being as in Fig.~\ref{f1}.  %The evolution of the field for different choices of mass of top in the region near to metastability.
}
\label{f_evol_metast}
\end{figure}

\begin{figure}[t]
\centering
\includegraphics[width=80mm]{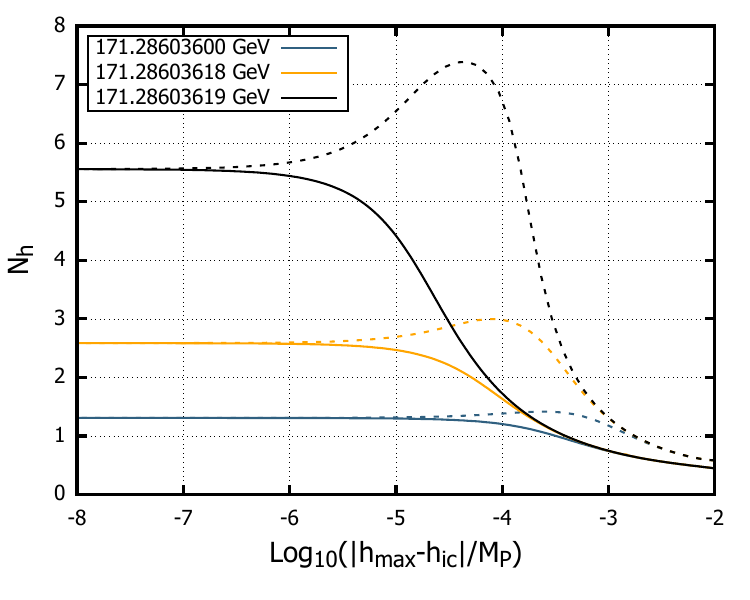}
\caption{Number of e-folds realised during the BEH spectator field domination phase, as a function of the difference between the initial BEH field value $h_{\rm ic}$ and the near-inflation point $h_{\rm cr}$ of the potential, for top quark masses corresponding to the stable regime, with the BEH mass and the strong coupling constant being as in Fig.~\ref{f1}.  The solid lines are for initial values lower than the inflexion point, $h_{\rm ic}<h_{\rm cr}$, while the dashed lines are for larger initial values, $h_{\rm ic}>h_{\rm cr}$.  }
\label{fig:field}
\label{res_back_inflection_matter}
\end{figure}

In Figs.~\ref{f_evol_metast} and \ref{res_back_inflection_matter} we depict the number of e-folds $\mathrm{N_{h}}$ that are realized in this phase for various initial field values, after solving  Eq.(\ref{eq:field}) for several choices of the top quark mass. 
%of top  using $\delta \mathrm{N}$ 
The range of initial conditions is chosen near the local maximum of the potential for top quark mass values in the metastability regime shown in Fig.~\ref{f_evol_metast}, or near the inflection point for values in the stability regime, extending down to values close to the bottom of the potential in both cases. %In particular, in  Fig.\ref{f_evol_metast}  we show the plot of $\mathrm{N_{h}}$ versus the value $\mathrm{\log_{10}(h_{max}-h_{ic})}$. The value $\mathrm{N_{h}}$ corresponds to the value of efolds when the integration of Eq.(\ref{eq:field}) stops.  
%More precisely, Fig.~\ref{f_evol_metast} focuses on initial conditions below the inlexion point while 
In Fig.~\ref{res_back_inflection_matter} we also show initial conditions above the near inflexion point (dashed lines).   In all cases, we find that more than one e-fold can be realized if the top-quark mass is sufficiently fine-tuned.  When this is translated in a curvature fluctuation, one therefore expects to generate PBHs.

%______________________________________________
%_______________________________________________

%\textbf{Primordial Black Hole formation -- }

The probability that a field fluctuation $\delta h_{\rm in}$, described by the stochastic dynamics during inflation, becomes a curvature fluctuation $\zeta \equiv \zeta_{\rm in} - \zeta_{\rm out}$  
%[3G] 
\textcolor{black}{(calculated as $\zeta_{\rm in} (x) =  N (\delta h_{\rm in} + \delta h_{\rm out} + \langle h \rangle ) - \langle N\rangle$) and $\zeta_{\rm out} (x) =  N ( \delta h_{\rm out} + \langle h \rangle ) - \langle N\rangle$)}, as studied in Ref.\cite{Stamou:2023vft,Carr:2019hud}, \textcolor{black}{leading to}
\be
P(\zeta) = \int {\rm d} \delta h_{\rm out} P(\delta h_{\rm in})  P(\delta h_{\rm out}) \left. \frac{{\rm d} h}{{\rm d} N}\right|_{h_{\rm out}} ~,
%\left. \frac{{\rm d} \psi}{{\rm d} N}\right|_{\langle \psi \rangle}~.
\label{eq:probability}
\ee
%
%\begin{equation}
%\begin{split}
%P=&P_1 \times P_2\\
%P_1 =& \frac{1}{\sqrt{2\pi \langle \delta h_{\text{1}}^2 \rangle}} \exp\left[ \frac{-(h - \langle h \rangle)^2}{2\langle \delta h_{\text{1}}^2 \rangle}\right]\\
%P_2 =& \frac{1}{\sqrt{2\pi \langle \delta h_{\text{2}}^2 \rangle}} \exp\left[ \frac{-(\delta h - h)^2}{2\langle \delta h_{\text{2}}^2 
%
%\begin{equation}
%P_4 = \frac{d\delta h(\delta N)}{d\delta N}
%\end{equation}
%
where $P(\delta h_{\rm in})$ and $P(\delta h_{\rm out}) $ are the Gaussian probability distributions of the corresponding field fluctuations during inflation, with variances $\mathrm{\langle \delta {h_{\text{in}}^2}\rangle}$ and $\mathrm{\langle \delta {h_{\text{out}}^2} \rangle}$  that can be approximated by Eqs.~\ref{eq:field1} and~\ref{eq:field2}. 

\begin{figure}[t]
\centering
\includegraphics[width=90mm]{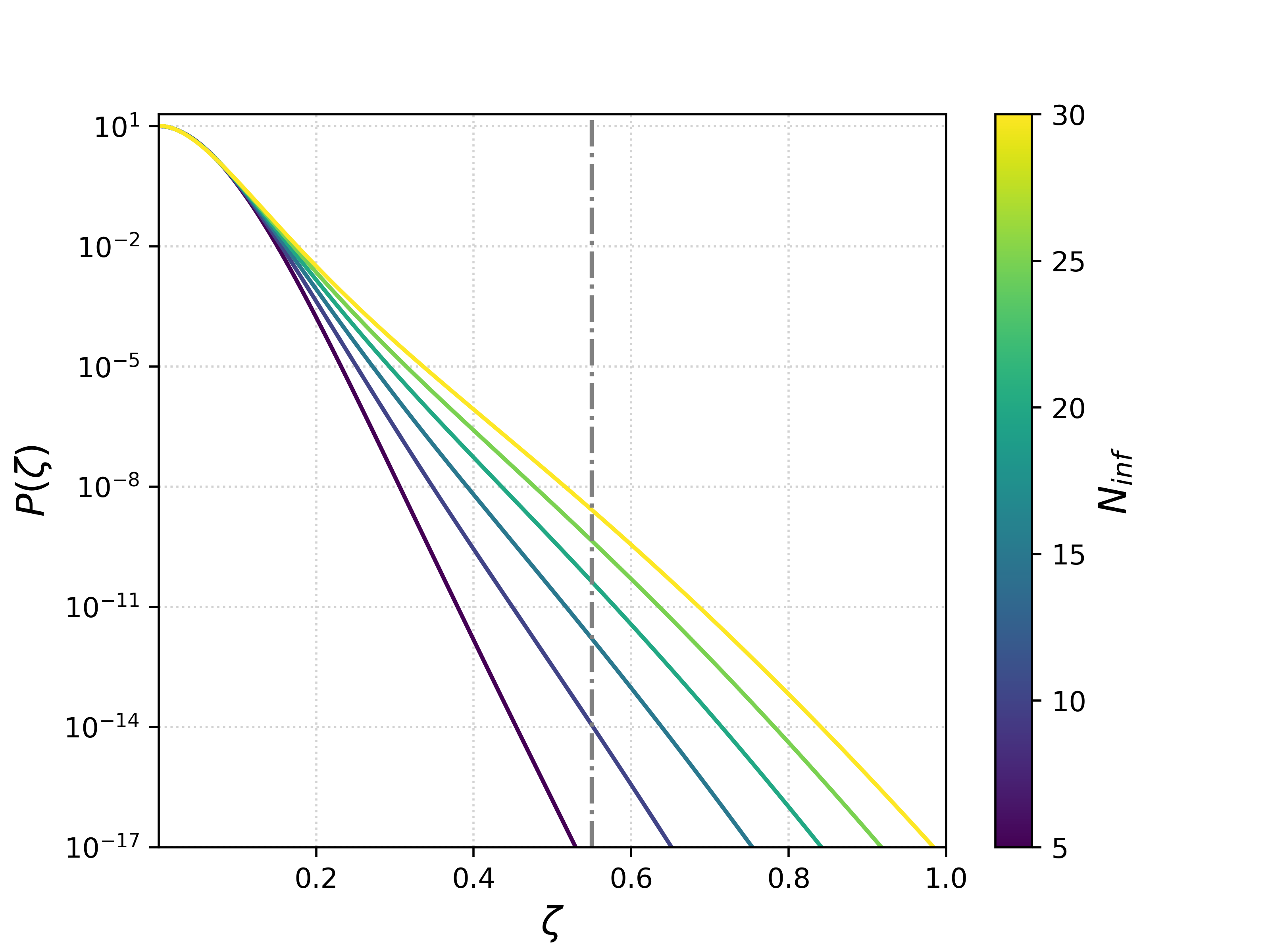}
\caption{Probability distribution of curvature fluctuations $\zeta$ produced by the SM BEH field, assuming  $\mathrm{m_{\rm t}}=$ 171.2860319 GeV, with $m_{\rm h}$ and $\alpha_{\rm s} $ as in Fig.~\ref{f1}.
%in respect to $\zeta$.  
The vertical line corresponds to the critical threshold $\zeta_{\rm cr}$ for PBH formation.  } 
\label{fig:prob}
\end{figure}

%$\langle N \rangle=4$

In Fig~\ref{fig:prob} we depict the probability distribution of $\zeta$, for different fluctuation sizes that are associated to different values of $N_{\rm inf}$.   Clearly, one can notice that these distributions, even if they become non-Gaussian as $N_{\rm inf}$ increases, always have a variance that is much larger for the observed $10^{-5}$ curvature fluctuations on CMB scales.  More precisely, the condition on the power spectrum of curvature fluctuations produced by the BEH field, in the $\delta N$ formalism, can be obtained as 
\begin{equation}
    \mathcal P_\zeta \simeq \frac{H^2_{*}}{4\pi^2 M_{\rm P}^2 }\left( \frac{{\rm d} N}{{\rm d}h}\right)^2 \Big |_{h=\langle h \rangle} \ll 2.1 \times 10^{-9}
    \label{eq:pr_cmb}
\end{equation}
whereas in our model this condition is not satisfied, because ${\rm d}N / {\rm d} h $ is at best of order $10^{3} M_{\rm Pl}$.%\textcolor{ao}{[3C] We remark here that we can obtain similar probability distributions for the other choices of the mass of top depicted in Fig.\ref{fig:field}. }

%Green line shows  the corresponding case of the Gaussian distribution. 
As $N_{\rm inf}$ increases, the distribution of curvature fluctuations becomes broader.  This allows the distribution to reach values larger than the threshold for PBH formation (gray dot-dashed line). However, despite the elegance of this mechanism, the large-scale curvature fluctuations do not pass the constraints imposed by CMB observations.   The origin of this problem comes from the fact that the field value at the inflection point or at the potential maximum is much larger than $H_{\rm inf}$ that characterizes the size of the quantum field fluctuations.  If future refined calculations of the effective BEH potential in the SM or some of its extensions (e.g. in supersymmetry or supergravity frameworks) show that the BEH potential exhibits an inflexion point pushed down to a value $h_{\rm cr} \lesssim 10^{-4} M_{\rm Pl}$, then we could select values of $\langle h \rangle$ close to the bottom of the potential that lead to tiny curvature fluctuations in agreement with CMB observations, but with a growing variance that could allow to cross the threshold of PBH formation on small scales, following the mechanism investigated in~\cite{Stamou:2023vft} for a generic spectator field.  Below, we explore another option with the Palatini formulation of Gravity, where there is no inflexion point but a plateau potential.

%______________________________________________
%_______________________________________________
%\textbf{Retrieve the CMB power spectrum}
\textbf{The Palatini Formulation:  one way to form PBHs and evade the CMB constraints --} 
We have highlighted the 
%intrinsic challenge posed by the 
impossibility for the BEH spectator field scenario to 
%reproduce the primordial power spectrum on CMB scales.  While reducing the top quark mass holds the promise to 
produce small enough curvature perturbations on CMB scales and at the same time lead to PBHs.
%.

\begin{figure}[t]
\centering
\includegraphics[width=80mm]{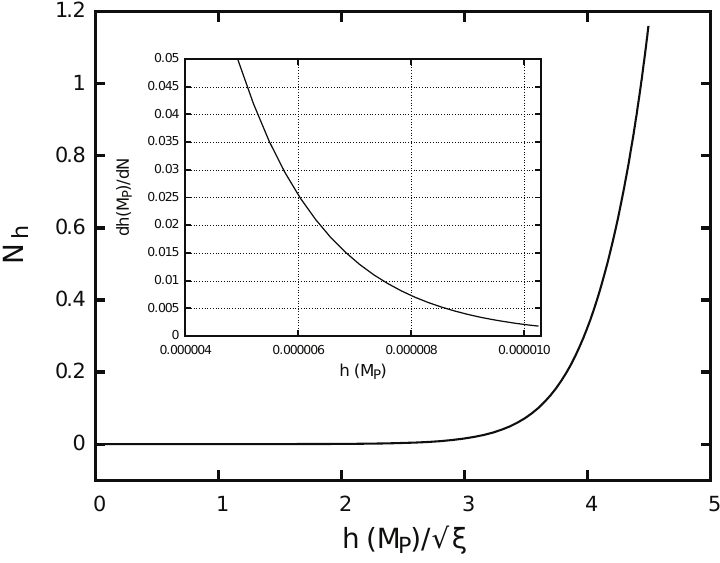}
\caption{Number of e-folds realised during the spectator field domination phase, as a function of the initial BEH field, assuming $\xi=4\times 10^{12}$.  The zoom plot shows the derivative of the field with respect to this number of e-folds.  }%The evolution of the field with respect to the number of efolds assuming $\xi=4\times 10^{12}$. In zoom plot is the derivative of the field with respect to the field.} 
\label{fig:retrieve}
\end{figure}

\begin{figure}[h!]
\centering
\includegraphics[width=100mm]{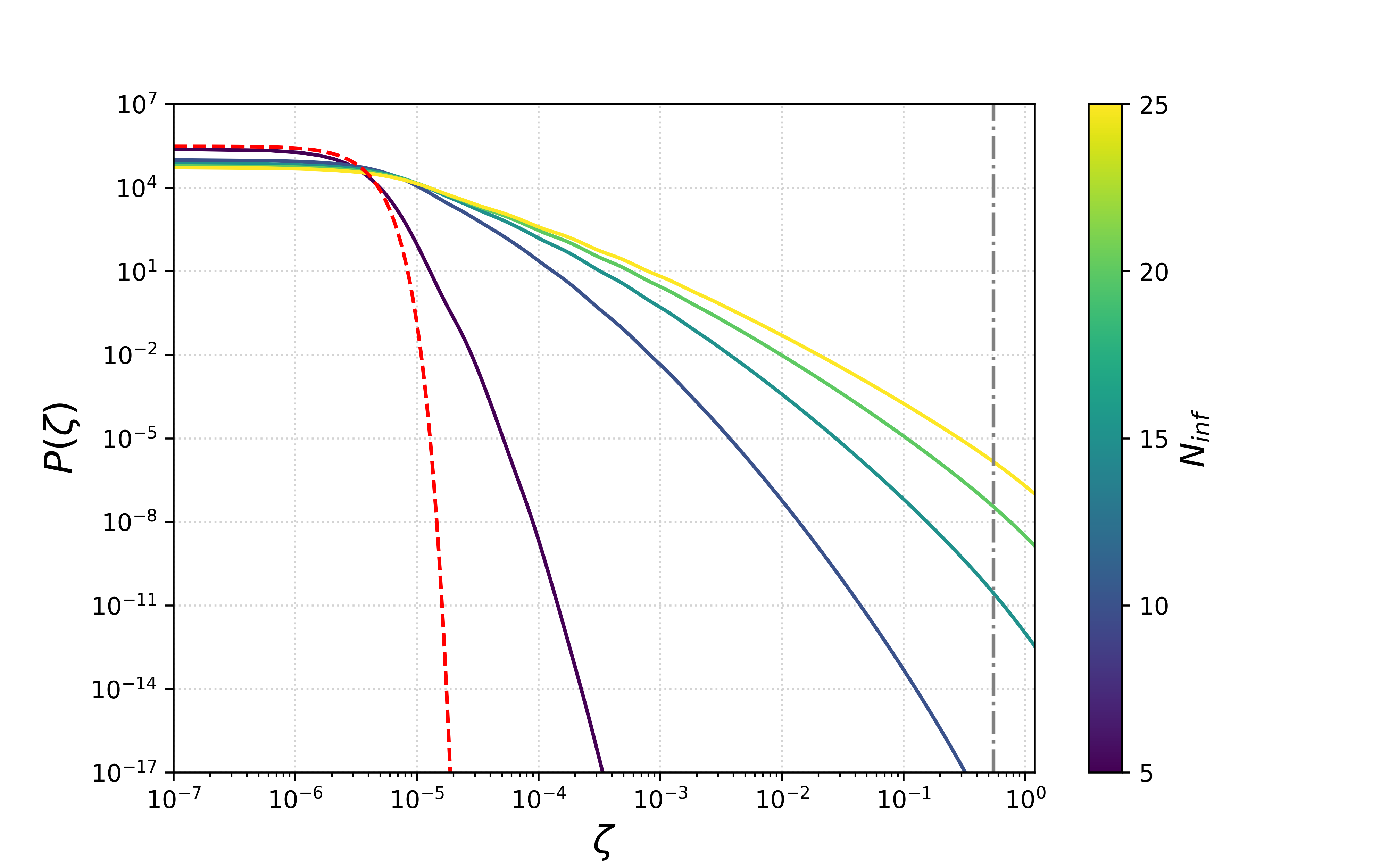}
\caption{Probability distribution of curvature fluctuations in the BEH-Palatini scenario, obtained at different scales related to the e-fold time $N_{\rm inf}$ during inflation at which it exits the Hubble radius.  The vertical line shows the considered curvature threshold $\zeta_{\rm cr}$ for PBH formation.} 
\label{fig:prob_pal}
\end{figure}

The BEH field could be the inflaton if it has a non-minimal coupling in the action \cite{Horn:2020wif, Rubio:2018ogq}.  A more general approach can be realized through the so-called Palatini formulation~\cite{Bauer:2008zj}, in which the connection $\Gamma^{\alpha}_{\beta \gamma}$ and the metric $\mathrm{g_{\mu\nu}}$ are treated as independent variables.  We now consider the case where the BEH field is not the inflaton but a spectator field and show that the changes in the effective potential can lead to the PBH production, while respecting CMB the constraints.   The Einstein-frame action is given by
\begin{equation}
\begin{split}
    S=&-\frac{1}{2}\int d^4 x \sqrt{- {g}} \big[{R}+K(h){g}^{\mu\nu}\partial_{\mu} h \partial_{\nu} h +\\ &+2 \Omega^{-4}(h)V(h)\big]
    \end{split}
\end{equation}
where the kinetic term is:
\begin{equation}
    K(h)=\frac{1}{ \Omega(h)^2} 
\end{equation}
and with
\begin{equation}
    \Omega^2=1+\xi h^2~,
\end{equation}
where $\xi$ is the non-minimal coupling of the field.  The potential with a fixed non-canonical kinetic term is given by \cite{Bauer:2008zj}
\begin{equation}
    U({\color{black}{\tilde h}})=\frac{1}{4{\xi^2}}{\lambda}\tanh^4\left( \sqrt{\xi} {\color{black}{\tilde h}} \right).
    \label{eq:palatini_pot}
\end{equation}
%[5]
\textcolor{black}{ The $\rm{\xi}$ parameter
%in both in tanh and out 
can be fixed such that curvature fluctuations from $\rm{\tilde h}$ are below the CMB level on cosmological scale. }
%ensure that we can fix the scales less than those of inflation and we can have a plateau very close to the origin in  order to respect the CMB. }
The potential has a plateau at large values of the field and could lead to inflation 
%\textcolor{black}{with $\rm{\xi < 1}$ }
\cite{Bauer:2008zj,Poisson:2023tja}.  In our case, we instead consider the regime $\xi\gg 1$.   
%
% assuming the  values for $\mathrm{m_t=171.286036}$ GeV, $\mathrm{m_h=125.7 }$ GeV and for the strong coupling $a_s=0.1184$, we find that the feature of inflection point is  no longer exist for $\xi\gg 1$.
% determinant of tanh and as a multiple term in the potential.    
The double advantage of this potential is that i) one can assume $\langle h \rangle \ll \xi^{-1} $ close to the bottom of the potential where the induced curvature fluctuations remains smaller than the ones observed on CMB scales, and ii) its stochastic fluctuations during inflation can reach the plateau region in rare regions that will generate large curvature fluctuations leading to PBHs. 
We solve the field dynamics for this potential and multiple initial field values, and depict in Fig.\ref{fig:retrieve} the expected additional expansion coming from the spectator field dominated phase.  In the inner panel, we show its variations with respect to the initial field value that are needed in the $\delta N$ formalism to compute the spectrum of curvature fluctuations.  In order to get $\mathcal P_{\mathcal R} < 2 \times 10^{-9}  $, according to Eq.~\ref{eq:pr_cmb}, one needs $ {\rm d} N / {\rm d} h_{\rm ic} < 50 $.  This condition is found to be satisfied when $\langle h \rangle < 8 \times 10^{-6} \Mpl$.  We then compute the probability of PBH formation in this model, following the methodology introduced above.  An example of the distribution of curvature fluctuations that is displayed on Fig.\ref{fig:prob_pal}.  It shows that a small fraction of them exceed the threshold for PBH formation.    %\lessim
Therefore, in the case of Palatini Formulation we can fulfill both the condition of small curvature fluctuations on CMB scales and get PBH formation on small scales.  

% 

%______________________________________________
%_______________________________________________
\textbf{PBH mass distribution -- }
PBHs formation is contingent upon surpassing a specific threshold for the density perturbations of the early universe. This threshold depends on various factors, including the equation of state of the primordial plasma~\cite{Musco:2023dak,Franciolini:2022tfm,Escriva:2022bwe,Atal:2019erb}. Additionally, the shape of the density perturbations plays a crucial role. In this section, we estimate the fractional abundances of PBHs for the SM BEH potential and for the potential obtained by using the Palatini formulation of Gravity. %[10]
\textcolor{black}{ One can relate the PBH mass $M_{\rm PBH}$ to the corresponding comoving fluctuation scale $k$ and to the e-fold time $N_{\rm inf}$ at which it exits the horizon during inflation through $M_{\rm PBH} /M_\odot \simeq [k/(2\times 10^7) {\rm Mpc}^{-1}]^{-2} \simeq 2 \times 10^{13} \exp(-2 N_{\rm inf}) $. }
%We remark here that we have the following characteristic wavenumber: the observable Universe's scale $(k_{H_0} \approx 2.3 \times 10^{-4} \, \text{Mpc}^{-1}$), the CMB pivot scale ($k_* = 0.05 \, \text{Mpc}^{-1}$), and the scale associated with PBHs and the QCD phase transition ($k_{QCD} \approx 10^6 \, \text{Mpc}^{-1}$). Between the scale of the observable Universe and the scale at which PBHs exit the horizon, inflation extends for approximately 22 e-folds.}
The density fraction of PBHs per unit of logarithmic mass at formation can be evaluated as 
\begin{equation}
\beta(M_{\rm PBH}) \equiv \frac{1}{\rho_{\rm cr}}\frac{{\rm d} \rho_{\rm PBH} }{{\rm d} \ln M_{\rm PBH}} = \int_{\zeta_{\rm cr}}^{\infty}  P(\zeta) {\rm d} \zeta \,
%2 \int_{N_{\psi}^{min}}^{ N_{\psi}^{max}}d N_h\int_{\zeta_{cr}}^{N_{\psi}^{max}} d\delta  N  \ p (N_\psi,\delta N,N_{inf} 
\label{eq:beta_derivative}
\end{equation}
where $\rho_{\rm cr}$ is the critical density, $\zeta_{\rm{cr}}$ denotes the threshold of the curvature fluctuation for PBH formation, and $\rm{P(\zeta)}$ denotes the probability density function described in Eq.~(\ref{eq:probability}).  %  
\begin{figure}[t]
\centering
\includegraphics[width=80mm]{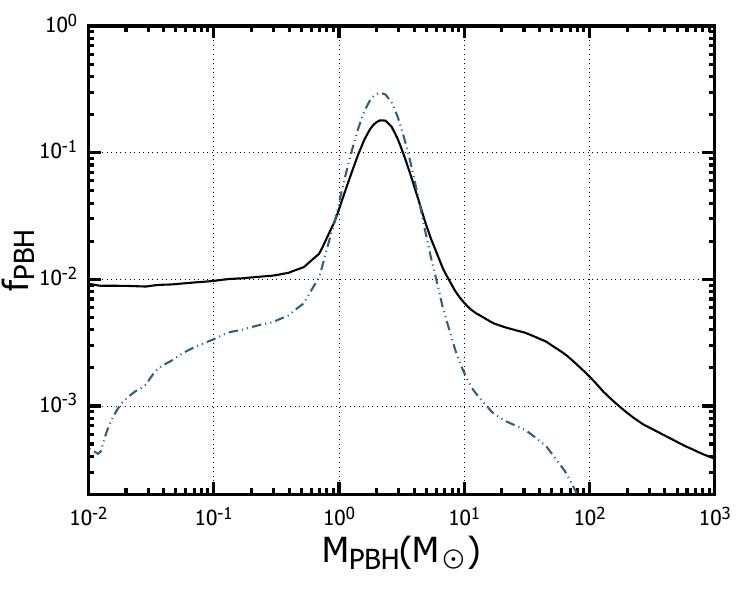}
\caption{Abundance of PBHs for the SM BEH field (without non-minimal coupling, 
%considering additional coupling constant 
solid line) for $\langle h \rangle= 0.185$ $\rm{M_P}$, and for the Palatini formulation of Gravity (dashed line) with   $\mathrm{\langle h \rangle=3.87\times 10^{-6}} \rm{M_P}$.  For the other parameters, we assume those of Figs. \ref{fig:prob} and \ref{fig:retrieve}.}
\label{f:fpbh}
\end{figure}
%[3C]
\textcolor{black}{In Eq.~\ref{eq:beta_derivative}, we implicitly assumed a top hat window function over unit intervals of $\ln k$ justified by one e-fold separation of the scales associated to   the inner and outer regions. }
For the threshold, we take into account its variations due to the QCD cross-over transition, following~\cite{Stamou:2023vft,Escriva:2022bwe}.  
%[34] 
\textcolor{black}{More precisely, we assumed that $\zeta_{\rm cr}=0.55$ in the radiation era and considered relative variations around this value at the QCD epoch.  One could question the validity of this threshold value.  In fact, there is so far no simulation of PBH formation in numerical relativity that assumes our probability distribution of curvature fluctuations, so the exact threshold value in the radiation era could be slightly different than the assumed one.  Nevertheless, we argue that the value of $\rm{\langle h \rangle}$ can alsways be anthropically selected to compensate a change in $\zeta_{\rm cr}$, in order to lead to the same PBH abundance.  Future work using dedicated numerical relativity simulations could help to better understand how the threshold vary is non-Gaussian models like ours.   } 
%Having evaluated the mass fraction of PBHs, 
Then we compute the today fractional abundances ${f_{\rm PBH}}$ as~\cite{Byrnes:2018clq}
\begin{equation}
f_{\rm PBH}(M_{\rm PBH}) \approx 2.4 \beta(M_{\rm PBH}) \left( \frac{2.8 \times 10^{17} M_\odot}{M_{\rm PBH}} \right)^{1/2}~. 
\end{equation}

The resulting abundance of PBHs %, denoted as $\mathrm{f_{PBH}}$
%, for both the Higgs potential solely and within the Palatini formulation 
is shown in Fig.\ref{f:fpbh}.  In both models, the abundance comes from the extra expansion after inflation and one can also notice the imprints from the QCD transition, with a peak at around $M_{\rm PBH} \approx 2 M_\odot$ and a bump from $30 M_\odot$ to $100 M_\odot$.  The value of $\langle h \rangle$ is adequately chosen to get all the dark matter made of PBHs, after integrating the mass distribution.  It is worth noticing that a different value can lead to lower dark matter fractions, without significantly changing the normalized mass distribution.  Given that it is a stochastic variable, with different realisations in different Universe patches, the total abundance of PBHs would in fact be typically lower or higher in other regions outside our observable Universe.  Therefore in this scenario one can invoke an anthropic selection argument to evade the usual fine-tuning problem linked to PBH formation~\cite{Carr:2019hud}. 

%[11]
\textcolor{black}{ The exact shape of the mass distribution also depends on the choice of slow-roll inflationary parameters.  An overproduction of light PBHs can be avoided even if the value of $\langle \delta h_{\rm out}^2 \rangle$ grows until the end of inflation, which tends to ease the formation of light PBHs.  The reason is that at the same time, $\langle \delta h_{\rm in}^2 \rangle$ is progressively suppressed when $N_{\rm inf} \epsilon_{2*} \gtrsim \mathcal O(1)$, i.e. when $N_{\rm inf} \gtrsim 25$.  For light PBHs, this latter effect  counterbalances the former.  Its importance depends on $\epsilon_{1*}$ and $\epsilon_{2*}$ and it influences the global shape of the distribution.}

Those results are consistent with those obtained in \cite{Stamou:2023vft} for a generic spectator field with a plateau-like potential.  Interestingly, there is no overproduction of light or supermassive PBHs and most of the abundance comes from stellar-mass PBHs.   For the mass distribution obtained in the Palatini framework, we even obtain a larger suppression of the mass function that is not reminiscent to the one obtained with a nearly scale invariant spectrum of Gaussian fluctuations as in~\cite{Carr:2019kxo}.  This type of mass functions could more easily evade microlensing and CMB constraints and have specific signatures in the PBH merging rates that could be observed with GW observations. 

We did not explore the impact of the radial profile of the curvature fluctuations or of the scaling relation of the critical threshold, which can alter the QCD features in the final PBH mass distribution, as well as the overall abundance since it can anyway be rescaled by an adequate choice of $\langle h\rangle$.

%______________________________________________
%_______________________________________________ 

%______________________________________________
%_______________________________________________

\textit{\bf Discussion and conclusion -- } Exploring the role of spectator fields offers a promising avenue  to understand the formation of PBHs, especially because of important fine-tuning issues in the vast majority of formation scenarios.   Unlike alternative models that rely on parameter fine-tuning, specific inflationary features or exotic physics, the generation of PBHs through the stochastic fluctuations of a light scalar field during inflation emerges as a relatively generic process. %
Moreover, the stringent observational constraints on the abundance and mass distribution of PBHs, as well as some possible positive observational evidences,  provide valuable insights into the early universe and the enigmatic properties of DM.  

In this study, by considering the BEH field as the most natural candidate for a light quantum stochastic spectator field during inflation, we have explored the ultimate possibility that PBHs could even have formed within the SM.  
%t
% reducing the need for excessive fine-tuning of model parameters.
However, our analysis reveals that while the BEH field can give rise to PBHs for specific values of strong coupling constant, of the top quark and of the BEH mass, leading to an inflexion point in the potential obtained by including radiative corrections, this would lead at the same time to excessive curvature fluctuations on cosmological scales compared to CMB anisotropy observations. As a result, this scenario appears to be excluded. % by current CMB data. 
However, we have also discussed the conditions under which such a scenario could become viable, which could lead to interesting connections between PBHs and the intriguing fact that the SM exactly lies at the transition between metastability and stability.  Then we also studied a variation of the scenario that assumes a non-minimal coupling to gravity in the context of the Palatini theoretical framework.  In this case, we obtained a viable scenario of PBH formation with a broad mass distribution and a peak at the solar-mass scale that could be tested with GW observations.  This scenario is a first concrete realisation of the generic mechanism that we investigated in~\cite{Stamou:2023vft}.

Our findings emphasize the interest to pursue the exploration of possible PBH formation scenarios based on the BEH field, also encouraging forthcoming studies to better understand why the SM parameters are such that the BEH field exactly lies between the metastable and stable regimes.  Other perspectives include more accurate determination of the PBH mass function, comparison with the black hole population inferred form GW observations and the computation of the scalar-induced GW background from our fully non-Gaussian curvature perturbations, on frequencies relevant for the recent GW observation with pulsar timing arrays.

%______________________________________________
%_______________________________________________

\bibliographystyle{apsrev4-2}

\bibliography{bib.bib} 

%\newpage

%\appendix

%\input{supp}

\end{document}